\documentclass[prl,twocolumn,showpacs,amsmath,amssymb]{revtex4}
\def\etal{\textit{et al.}}
\def\ibid{\textit{ibid. }}

\usepackage{graphicx}
\usepackage{dcolumn}
\usepackage{bm}
\usepackage{epsfig}

\begin{document}


\title{Conduction in Carbon Nanotubes Through Metastable Resonant States}

\author{Zhengfan Zhang}
\author{Venkat Chandrasekhar}
\affiliation{%
Department of Physics and Astronomy, Northwestern University,
Evanston, Illinois 60208, USA
}%

\author{Dmitriy A. Dikin}%
\author{Rodney S. Ruoff}
\affiliation{%
Department of Mechanical Engineering, Northwestern University,
Evanston, Illinois 60208, USA
}%
\date{November 12, 2003}

\begin{abstract}
We report here on electrical measurements on individual
multi-walled carbon nanotubes (MWNTs) that show that the presence
or movement of impurities or defects in the carbon nanotube can
radically change its low temperature transport characteristics.
The low temperature conductance can either decrease monotonically
with decreasing temperature, or show a sudden increase at very low
temperatures, sometimes in the same sample at different times. This unusual behavior
of the temperature dependence of the conductance is correlated
with large variations in the differential conductance as a
function of the dc voltage across the wire. The effect is well
described as arising from quantum interference of conduction
channels corresponding to direct transmission through the nanotube
and resonant transmission through a discrete electron state, the
so-called Fano resonance.
\end{abstract}

\pacs{73.63.Fg, 85.35.Kt, 73.20.Hb}
\maketitle There is tremendous interest in the transport
properties of carbon nanotubes due to their potential for use in
future nanodevices \cite{nextgeneration}, and from their role as
canonical models of one-dimensional electron transport
\cite{luttinger}. While individual single-walled carbon nanotubes
are expected to show either semiconducting or metallic behavior
depending on their chirality \cite{physicalproperties}, the
presence of impurities, defects and interactions is expected to
modify this behavior \cite{backscattering,defects}. However,
measuring the intrinsic transport properties of single or
multiwalled carbon nanotubes is complicated by the experimental
problem of making electrical contact to the nanotube. Quite often,
the contact resistance between the metallic electrodes and
nanotube is much higher than the resistance of the nanotube
itself, so that the transport properties of the device are
determined in large part by the properties of the metal-nanotube
contacts. In spite of this problem, transport measurements on high
contact-resistance carbon nanotube devices by a number of groups
have elucidated the wide variety of physical problems that can be
studied in these systems, including the Coulomb blockade
\cite{set}, the Kondo effect \cite{kondo}, and Luttinger liquids
\cite{luttinger}.

In devices with lower contact resistances, the intrinsic
properties of carbon nanotubes can be directly measured. For
example, by using different techniques to obtain low resistance
contacts, observation of quantization of the conductance in
ballistic carbon nanotubes has been reported \cite{ballistic}, and
there are also suggestions of superconductivity in nanotube rope
devices made with low resistance contacts
\cite{superconductivity}. However, much interesting physics still
remains to be explored. In this Letter, we report on transport
measurements on multiwalled carbon nanotubes (MWNTs) with low
resistance contacts. The low-temperature differential conductance
$G(V_{dc})$ of these devices as a function of the dc voltage bias
$V_{dc}$ across them is highly asymmetric, showing large,
reproducible fluctuations that can be as large as 10\% of the
total conductance. This conductance `fingerprint' can change on
thermal cycling to just 2 K, indicating that it might be
associated with metastable impurities or mechanical instabilities.
The fluctuations in $G(V_{dc})$ are reflected in the zero-bias
($V_{dc}$=0) conductance as a function of temperature $G(T)$,
which can either decrease or increase with decreasing temperatures
at low temperatures, sometimes showing both dependences in a
single sample upon thermal cycling. The sharp structure in
$G(V_{dc})$ can be described as arising from Fano resonances
through resonant states in the device. The position of these peaks
and dips frequently change on thermal cycling, indicating that the
resonant states may arise from metastable impurities or defects in
the device.

\begin{figure}[hb]
\includegraphics[width=8cm]{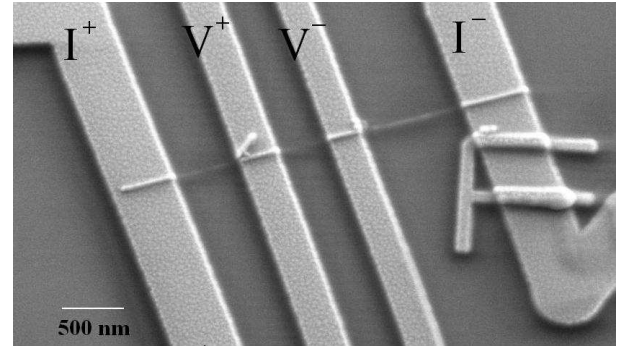}
\caption{Scanning electron microscope (SEM) image of Sample 3, a
single MWNT with four electrodes. The legends show the probe
configurations used for making the four terminal resistance
measurements. For this sample, each electrode had two external
contacts, so that the four-terminal resistance of the two inner
metal-nanotube contacts could be directly measured. } \label{fig1}
\end{figure}

Our devices consist of an isolated MWCNT with four Au/Ti
electrodes (Fig. 1). The arc-grown MWNTs, typically 2-5 $\mu$m
long and 25-50 nm wide, were spun onto an oxidized Si substrate
from a suspension in dimethylformamide. After locating the
nanotubes with respect to fiducial marks, and patterning
electrodes by electron-beam lithography, thin film electrodes (5
nm Ti/50 nm Au) were deposited to make contact to the nanotube.
Prior to deposition, a short-time oxygen plasma etch was used to
improve the metal-nanotube contact resistance. Without this
cleaning, the contact resistances were typically in the range of a
few k$\Omega$ to a few M$\Omega$, while this process reduced the
contact resistance to a few hundred ohms. The presence of four
electrodes enables us to make four-terminal resistance
measurements on the MWCNTs, eliminating effects of the contact
resistance between the MWCNT and the electrodes. In some devices,
two external contacts were made to each electrode, as shown in
Fig. 1, enabling direct measurements of the electrode-nanotube
contact resistance; in the other devices, the contact resistance
was inferred from differences between four-terminal and
two-terminal measurements. The typical distance between the
voltage probes was 300 nm. The samples were measured in a $^3$He
refrigerator and a dilution refrigerator. A home-made ac
resistance bridge was used to measure the four-terminal
differential resistance $dV/dI$ as a function dc current $I_{dc}$
using the probe configuration shown in Fig. 1, with ac excitation
in the range of 0.25-1 nA to avoid heating effects. In this paper,
we shall plot our measurements in terms of the differential
conductance $G$=1/($dV/dI$) as a function of the dc voltage
$V_{dc}$, obtained by numerically integrating the measured $dV/dI$
vs $I_{dc}$ curves.

\begin{figure}[ht]
\includegraphics[width=8.7cm]{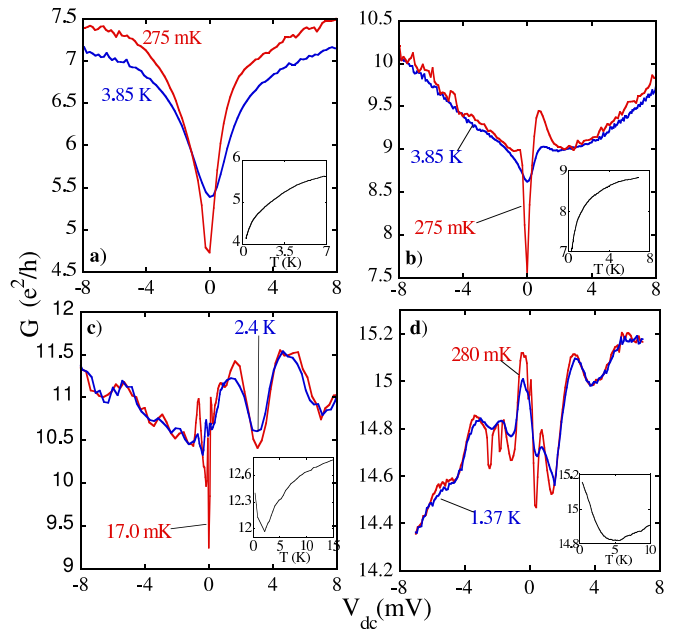}
\caption{$G(V_{dc})$ of 4 samples, in units of $e^2/h$. The insets
show $G(T)$. The average contact resistances (per contact) are a)
Sample 1: 1 k$\Omega$ at 4 K. b) Sample 2: 900 $\Omega$ at 4 K. c)
Sample 3: 133 $\Omega$ and 173 $\Omega$ at 12 K for the two inner
contacts. d) Sample 4: less than 100 $\Omega$ at 4 K. Contact
resistances were inferred from differences between 2- and
4-terminal measurements of the devices, except for Sample 3, where
they could be measured directly. } \label{fig2}
\end{figure}

Figure 2 shows $G(V_{dc})$ for four different samples with contact
resistances ranging from 1 k$\Omega$ to less than 100 $\Omega$, at
two different temperatures each, and demonstrates the
sample-specific behavior that can be seen in our devices.
$G(V_{dc})$ is highly asymmetric and shows large fluctuations,
with the pattern of the fluctuations being different for different
devices. In general, the structure in $G(V_{dc})$ becomes sharper
as the temperature is lowered: the peaks increase in conductance,
while the valleys decrease in conductance. This gives rise to
characteristically different temperature dependences for nominally
identical samples: for example, if a peak is observed at
$V_{dc}$=0, $G(T)$ will increase with decreasing temperature,
while if a valley is observed, it will decrease. (Of course, it is
not necessary that either a peak or a valley occur at $V_{dc}=0$).
This is demonstrated in the insets to the panels in Fig. 2, which
show the corresponding $G(T)$ for each device. As demonstrated in
Fig. 2, the structure in $G(V_{dc})$ appears to increase on
average as the resistance of the contacts decreases. Indeed, large
fluctuations in $G(V_{dc})$ are not seen in our samples with
contact resistances in the k$\Omega$ to M$\Omega$ range; one sees
instead a largely symmetric curve with reduced conductance near
$V_{dc}$=0, and correspondingly, a monotonically decreasing
$G(T)$.

\begin{figure}[hb]
\epsfig{file=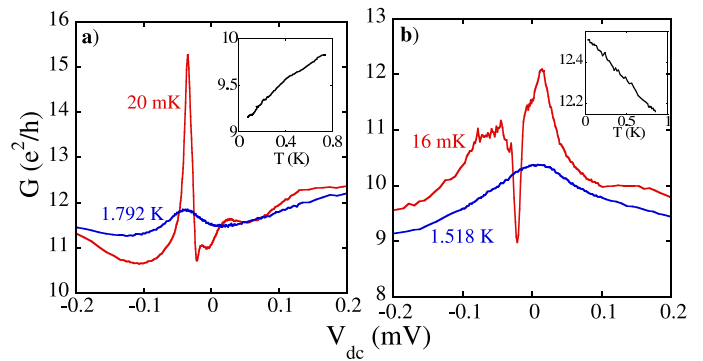, angle=-90, width=9cm} \caption{$G(V_{dc})$
of Sample 3 at two temperatures, on two different cooldowns of the
dilution refrigerator, in units of $e^2/h$. The insets show the
measured temperature dependence of $G(T)$. a) First cooldown. b)
Second cooldown. Note that $G(T)$ does not match $G(V_{dc}=0)$
exactly, as the sample characteristics changed even on warming to
2 K.} \label{fig3}
\end{figure}

$G(V_{dc})$ varies not only from sample to sample, but can also
change in a single sample as a function of thermal cycling. Figures
3(a) and 3(b) show $G(V_{dc})$ for Sample 3 (see the caption of
Fig. 2), for two different cooldowns of the dilution refrigerator.
As can be seen, there is a radical change after warming to room
temperature: while the most prominent feature at low temperature
in Fig. 3(a) is a peak, this feature has changed into a dip in
Fig. 3(b). As we noted before, this change is also reflected in
$G(T)$, as shown in the inset to the panels in Fig. 3. Indeed, we
found that it was not necessary to warm the samples all the way to
room temperature; warming to temperatures on the order of a few
Kelvin changed the behavior of the sample, although the change was
not as dramatic as shown in Figs. 3(a) and 3(b).

Sample-specific fluctuations of the conductance are well known in
mesoscopic systems. In the case of disordered metals and
semiconductors, they are associated with quantum interference of
electron waves that are scattered by impurities, defects or grain
boundaries in the sample \cite{ucf}. Each scattering event
introduces a finite but time-independent shift in the phase of the
electron. The phases and phase shifts of the electrons can be
modulated by external parameters such as a gate voltage or
magnetic field. So long as time-dependent scattering processes
that destroy the phase of the electron (processes such as
electron-electron or electron-phonon scattering) are negligible,
the interference of electrons manifests itself as aperiodic or
periodic fluctuations of the conductance as a function of the
external parameter \cite{ucfexp}. Furthermore, since scattering
from each impurity or defect introduces a phase shift, the
movement of the impurity by even a very small distance (equivalent
to 1/$k_F$, where $k_F$ is the Fermi wave vector) will change the
phase shift, resulting in a corresponding change in the
conductance pattern \cite{noise}. Time-dependent conductance
changes have been observed in disordered metals, and this
mechanism has been shown to be a source of $1/f$ noise in metals
at low temperatures \cite{noiseexp}.

A similar interference mechanism can also exist in relatively
clean carbon nanotubes with one or a few number of impurities or
defects. In this case, the interference can be between electron
waves that are directly transmitted, and those that are
transmitted via a resonant state. This interference between
directly transmitted channels and resonant channels gives rise to
a Fano resonance, well known in atomic scattering \cite{Fano}. For
simplicity, we consider the case of a single directly transmitted
channel and a single resonant state. Transmission through the
resonant channel is described by an amplitude
$t_r(\epsilon)=z_r\Gamma/(2(\epsilon-\epsilon_0)+i\Gamma)$
\cite{Clerk}. Without the factor $z_r$, this gives the usual
expression for resonant transmission through a localized state of
energy $\epsilon_0$ (measured with respect to the Fermi energy
$\epsilon_F$) and intrinsic energy width $\Gamma$,
$T_r=|t_r|^2=\Gamma^2/(4(\epsilon-\epsilon_0)^2+\Gamma^2)$, with a
transmission of $T_r=1$ on resonance ($\epsilon=\epsilon_0$). The
transmission amplitude of the direct channel does not depend on
$\epsilon$, and can be described by an expression of the form
$t_d=\sqrt{T_d}e^{i\alpha_d}$, where $T_d$ is the transmission of
the direct channel, and $\alpha_d$ describes the phase difference
between the direct and resonant path. The interference between the
two paths is taken into account by taking the sum of the
transmission amplitudes to calculate the total transmission
coefficient, $T_t=|t_r+t_d|^2$. The resulting conductance can be
expressed in the Fano form
\begin{eqnarray}
G(\epsilon)=\frac{2e^2}{h}T_t=\frac{2e^2}{h}T_d\frac{|2(\epsilon-\epsilon_0)+q\Gamma|^2}{4(\epsilon-\epsilon_0)^2+\Gamma^2}
\end{eqnarray}
where $q=i+z_re^{-i\alpha_d}/\sqrt{T_d}$ is the complex Fano
parameter \cite{Clerk}. Note that far from resonance ($\epsilon
\gg \Gamma$), the conductance reduces to $G=(2e^2/h) T_d$.

A number of characteristics distinguish the resulting Fano
resonance from other resonances that might occur in carbon
nanotubes. First, depending on the phase difference between the
resonant and non-resonant transmission channels, the Fano
resonance can give rise to a peak or dip in the conductance (or
something in between). Second, the Fano lineshape can be
asymmetric about the transmission maximum or minimum. Finally, the
position of the resonance is determined by the energy of the
resonant state, and does not necessarily occur at the Fermi energy
(zero bias). These distinguishing characteristics, which can be
clearly seen in our data, rule out other possible mechanisms (such
as the Kondo effect \cite{kondo}) for the structure we observe in
the differential resistance of our devices.

Two groups have recently reported observing Fano resonances in
carbon nanotube devices. Kim \textit{et al.} \cite{Kim} measured the
conductance of crossed MWNTs, and observed a Fano resonance in two
of the devices. Noting that a Fano resonance was never observed in
devices without crossed MWNTs, they associated the presence of the
Fano resonance with the MWNT cross, although the mechanism by
which a discrete electron level is created was not discussed. Yi
\textit{et al.} \cite{Yi} measured the conductance of crosses consisting of
metal electrodes patterned across MWNT bundles. In two of these
devices, they observed non-monotonic behavior of the conductance
near zero voltage bias, which they ascribed to a Fano resonance
arising from interference between a Kondo resonance and
non-resonant channels. However, the nature of the localized state
giving rise to the Kondo resonance was not made clear.
Furthermore, this interpretation is suspect in our opinion,
because a Kondo resonance typically arises at the Fermi energy,
and the resonances observed by Yi \textit{et al.} were typically observed at
voltage biases of ~0.3-0.6 mV. In both papers, the metastable
behavior we observe was not reported.

\begin{figure}[ht]
\includegraphics[width=8.7cm]{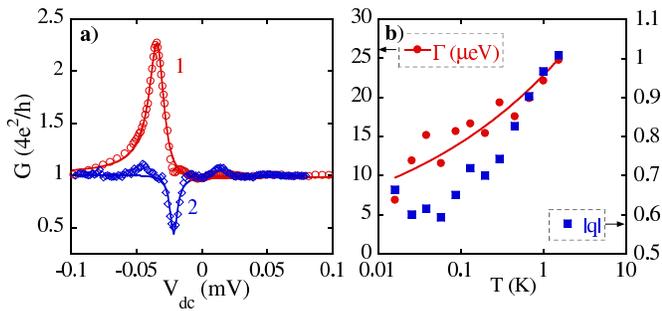}
\caption{a) The Fano resonances in the conductance of Sample 3
from Fig. 3(a) at $T$=20 mK (curve 1), and Fig. 3(b) at 16 mK
(curve 2). The data (open circles) are normalized to $4e^2/h$, as
discussed in the text. The solid lines are fits to the Fano
function. b) the temperature dependence of the fitting parameters
$\Gamma$ and $|q|$ of the data from Fig. 3(b), on a semilog plot.
The solid line is a power-law fit to $\Gamma$= 22.9 x $T^{0.21}$ $\mu$eV.
The phase of $q$ is nearly $\pi/2$ ($1.58 \pm 0.06$). }
\label{fig4}
\end{figure}

Fig. 4(a) shows the measured Fano resonances for the lower
temperatures shown in Figs. 3(a) and 3(b), along with fits to the
Fano function given above. To obtain these curves, the background conductance  outside the region $V_{dc}$
= [-0.1 mV,-0.02 mV] for curve 1, and $V_{dc}$ = [-0.05 mV, 0.02
mV] for curve 2 has been fit, and subtracted from the experimental data. After
subtraction, the conductance far from resonance would be 0.
However, we arbitrarily introduce an offset of $4 e^2/h$,
corresponding to the conductance of two channels of
a single-walled nanotube that coherently interfere with the
resonant state, with the assumption that other channels that
contribute to the background do not interfere with the resonant
state. It can be seen that data are well described by the Fano
equation. For curve 1, $t_d$ is in phase with $t_r$ on resonance,
resulting in a maximum of conductance, while for curve 2, $t_d$ is
out of phase with $t_r$ on resonance, resulting in a minimum of
conductance. This change in the phase of the non-resonant channel
and the resonant channel is a result of the annealing process,
which presumably causes a change in the position of the impurity
or defect that gives rise to the resonant state.

Fig. 4(b) shows the fitted value of $|q|$ and $\Gamma$ as a
function of temperature for the conductance dip shown in Fig.
4(a). Both $|q|$ and $\Gamma$ increase with increasing
temperature. Although the values we obtain for $\Gamma$ are
comparable to those obtained by Kim \etal, the temperature
dependence is different. Kim \textit{et al.} observed a linear temperature
dependence, which they ascribed to thermal broadening of the
linewidth, even though $\Gamma$ was less than $k_B T$. Our
temperature dependence is not linear, which is not surprising,
since we are in the regime $\Gamma \le k_B T$. Fitting to a power
law gives a dependence $\Gamma \simeq T^{1/5}$, shown as the solid
line in Fig. 4(b); the origin of this power law is not clear to
us. In contrast, while $|q|$ also increases with temperature at
higher temperatures, it appears to saturate below 100 mK.   The
phase of $|q|$ remains essentially constant over the fitted
temperature range.

What is the possible origin of the resonant states in our samples?
Although we are not certain, the metastability we observe suggests
impurities or structural defects cause variations in the potential
seen by the electrons in the nanotube, which in turn lead to
localized electronic states with well-defined resonant energies.
Movement of the impurities or defects can give rise changes in the
conductance through interference effects. Although the changes we
observed in our samples typically occurred on the scale of days at
low temperatures, we have also observed changes on the scale of
hours. If there are many such impurities moving on a sufficiently
rapid time scale, this may be one mechanism for the large $1/f$ noise
in carbon nanotubes reported by some groups \cite{tubenoise}.

\begin{acknowledgments}
We thank the group of R. P. H. Chang for providing us the
nanotubes used in these experiments, and D. E. Prober for a
critical reading of the manuscript. Funding for this work was
provided by a NASA/MSFC Phase II SBIR, Contract No.
NAS8-02102, through a subcontract from Lytec, LLC.
\end{acknowledgments}



\end{document}